\documentclass{article}
\usepackage{ijcai24}

\usepackage{times}
\usepackage{soul}
\usepackage{url}
\usepackage[hidelinks]{hyperref}
\usepackage[utf8]{inputenc}
\usepackage[small]{caption}
\usepackage{graphicx}
\usepackage{amsmath}
\usepackage{amsthm}
\usepackage{booktabs}
\usepackage{algorithm}
\usepackage{algorithmic}
\usepackage[switch]{lineno}

\usepackage{amsmath}
\usepackage{amssymb}
\usepackage{booktabs}
\usepackage{float}
\usepackage{pifont}
\usepackage{hyperref}

\title{Beyond Alignment: Blind Video Face Restoration via Parsing-Guided \\ 
Temporal-Coherent Transformer}

\author{
Kepeng Xu$^1$\and
Li Xu$^1$\and
Gang He\thanks{Corresponding Author}$^1$\and
Wenxin Yu$^2$\and
Yunsong Li$^1$\\
\affiliations
$^1$Xidian University,
$^2$Southwest University of Science and Technology\\
\emails
kepengxu11@gmail.com,
ghe@xidian.edu.cn
}

\begin{document}

\maketitle

\begin{abstract}
    Multiple complex degradations are coupled in low-quality video faces in the real world. 
    Therefore, blind video face restoration is a highly challenging ill-posed problem, 
    requiring not only hallucinating high-fidelity details but also enhancing temporal coherence across diverse pose variations. 
    Restoring each frame independently in a naive manner inevitably introduces temporal incoherence and artifacts from pose changes and keypoint localization errors. 
    To address this, we propose the first blind video face restoration approach with a novel parsing-guided temporal-coherent transformer (PGTFormer) without pre-alignment. PGTFormer leverages semantic parsing guidance to select optimal face priors for generating temporally coherent artifact-free results. 
    Specifically, we pre-train a temporal-spatial vector quantized auto-encoder on high-quality video face datasets to extract expressive context-rich priors. Then, the temporal parse-guided codebook predictor (TPCP) restores faces in different poses based on face parsing context cues without performing face pre-alignment. 
    This strategy reduces artifacts and mitigates jitter caused by cumulative errors from face pre-alignment. Finally, the temporal fidelity regulator (TFR) enhances fidelity through temporal feature interaction and improves video temporal consistency. 
    Extensive experiments on face videos show that our method outperforms previous face restoration baselines. The code will be released on \href{https://github.com/kepengxu/PGTFormer}{https://github.com/kepengxu/PGTFormer}
\end{abstract}

\section{Introduction}

Deep learning technology has developed rapidly recently. 
Researchers have expanded deep learning technology into different fields such as target recognition
\cite{zhang2023superyolo,zhang2023guided,wang2024patchhar,zhang2024efficientmfd}, 
natural language processing\cite{chen2024modeling,chen2022hierarchical,chen2020coarse}, 
and video enhancement\cite{xu2022transcoded,hdcfmv0,xu2023towards,itesss,xu2022sdrtv}, 
video restoration\cite{wang2024selfpromer,wang2024correlation,wang2024promptrestorer}, etc.
The human face, traditionally considered a focal point in the video, presents an intriguing opportunity — leveraging limited computational resources to enhance areas of interest (face area). Currently, researchers have pivoted towards real-world face restoration, conceptualizing the task as blind face restoration to restore high-quality faces from real face suffering from multiple degradations.

\begin{figure}[t]
\centering
\includegraphics[width=0.99\linewidth]{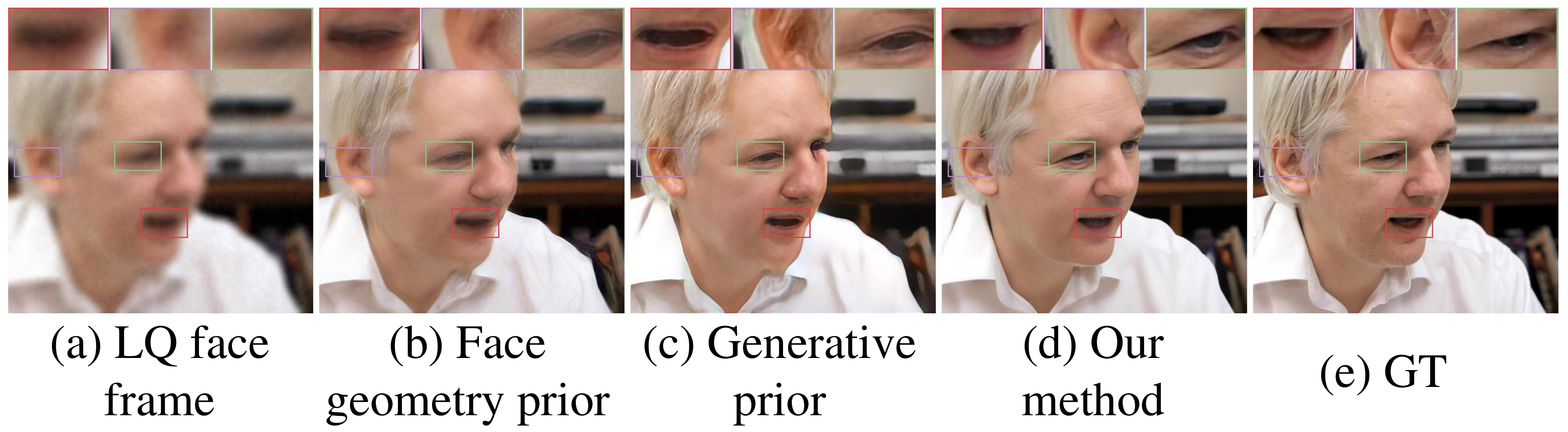} 
\caption{
(a) Low-quality face input need to be restored.
(b) Face geometry-based methods are able to perceive facial structure and can avoid wrong structures, but textures are blurry.
(c) The methods based on generative priors can generate detailed textures. However, due to the positional bias of the face structure, artifacts will occur when the input face is not a standard aligned face.
(d) Our method has the capability to yield more natural face components in regions prone to artifacts.
}
\label{parandvqgan}
\vspace{-0.35cm}
\end{figure}

The amalgamation of various degradations (including noise, blur, and coding artifacts) poses significant challenges in achieving high-quality face restoration. Earlier methodologies directly incorporated face geometry \cite{Chen_Li_Yang_Lin_Zhang_Wong_2021,Chen_Tai_Liu_Shen_Yang_2018} as a prior into the model. Despite their ability to perceive global structure, as depicted in Fig.\ref{parandvqgan} (b), these methods fall short in generating high-quality texture details.
Techniques \cite{Yang_Ren_Xie_Zhang_2021,Wang_Li_Zhang_Shan_2021} grounded in generative priors have shown potential in producing high-quality texture details. However, generative models suffer from fixed spatial bias, leading to various erroneous illusions and artifacts when processing low-quality faces with complex poses, as shown in Fig.\ref{parandvqgan} (c).

\begin{figure}[t]
\centering
\includegraphics[width=0.99\linewidth]{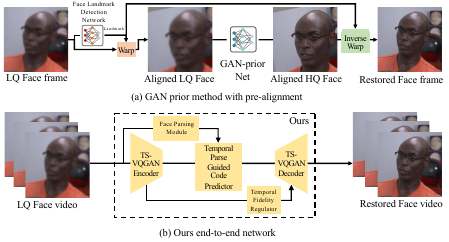} 
\caption{(a) GAN prior method with pre-alignment. In video face restoration, landmark detection on low-quality faces will produce errors, and inter-frame errors will lead to discontinuous restoration results.
(b) Our end-to-end framework. The proposed time-parsing guided Transformer does not require pre-alignment.}
\label{workflowprealigment}
\vspace{-0.2cm}
\end{figure}

Furthermore, previous works \cite{Wang_Li_Zhang_Shan_2021,Zhou_Chan_Li_Loy_2022,Gu_Wang_Xie_Dong_Li_Shan_Cheng} have primarily focused on image-based face restoration. However, in real-world scenarios, face restoration is not limited to images alone; it is also crucial in videos.
In the context of video face restoration, there are unique challenges and unresolved issues that need to be addressed. One such challenge is the variability in face pose across frames. When using existing prior generation methods, the workflow involves 1. Pre-aligning low-quality face images to standard poses. 2. Face restoration. 3. Reverse warping to the input face, as shown in the Fig.\ref{workflowprealigment} shown in (a). Unfortunately, this approach introduces additional degradation problems during frame-by-frame processing. These problems arise from errors in low-quality face keypoint detection, which in turn affect the accuracy of warping. Consequently, facial structure generation errors occur, and the accumulation of such errors can amplify inter-frame differences, leading to video jitter. These issues remain unsolved by prior methods relying on face-generation priors.

To tackle the aforementioned challenges, we propose a novel video face restoration approach to overcome the limitations of previous techniques by addressing the issues associated with changing face poses, error propagation, and video jitter. In contrast to previous image-based face restoration approaches, our proposed method aims to restore high-quality face texture details consistently across frames, while avoiding the generation of undesirable artifacts, hallucinations, and frame jitters.
To achieve video face restoration, we introduce a temporal-spatial interaction model that facilitates the enhancement of facial information. Moreover, we leverage a face parse-guided module to restore faces from videos without the need for pre-alignment.

Our approach introduces a Temporal-Spatial VQGAN (TS-VQGAN) based on the Swin3D \cite{liu2022video} Transformer, enabling the extraction of precise timing information from video frames to obtain an accurate representation of video face feature. The trained codebook and decoder encompass a rich set of high-quality video face priors.

Furthermore, we propose a codebook prediction mechanism guided by temporal context parsing. Specifically, we fuse the low-quality (LQ) features from consecutive frames and input them into a 3D Transformer, referred to as Temporal Parsing-guide Code predictor (TPCP), to enhance the accuracy of codebook predictions using contextual information from the video. Notably, we incorporate face parsing features as position encoding inputs to the 3D Transformer, addressing the issue of artifact hallucinations caused by fixed spatial biases in global face representations. By adopting this approach, we mitigate the generation of unnatural artifacts, even when dealing with complex face poses. This novel technique significantly expands the applicability of generative prior-based methods in video face restoration.

Finally, we introduce a novel module called the Temporal Fidelity Regulator (TFR), which plays a crucial role in integrating multi-scale low-quality features with the decoder stage. To ensure the temporal consistency of video faces, our TFR extracts temporally consistent face features, which are then utilized to modulate the HQ features. This enables the decoder to achieve coherent recovery of high-fidelity video faces, promoting visual consistency throughout the sequences.


\textbf{Contribution}: 1) Our work is pioneering as the first end-to-end method for video face restoration; 2) We introduce a tailored temporal-spatial interactive network structure that effectively utilizes multi-frame face information, yielding superior restoration results; 3) Furthermore, we present a novel approach for high-quality face restoration without pre-alignment, significantly broadening the scenarios of face restoration applications; 4) Our proposed scheme showcases state-of-the-art performance on video face restoration tasks, as validated through extensive experiments.

\begin{figure*}[h]
\vspace{-0.3cm}
\centering
\includegraphics[width=0.9\linewidth]{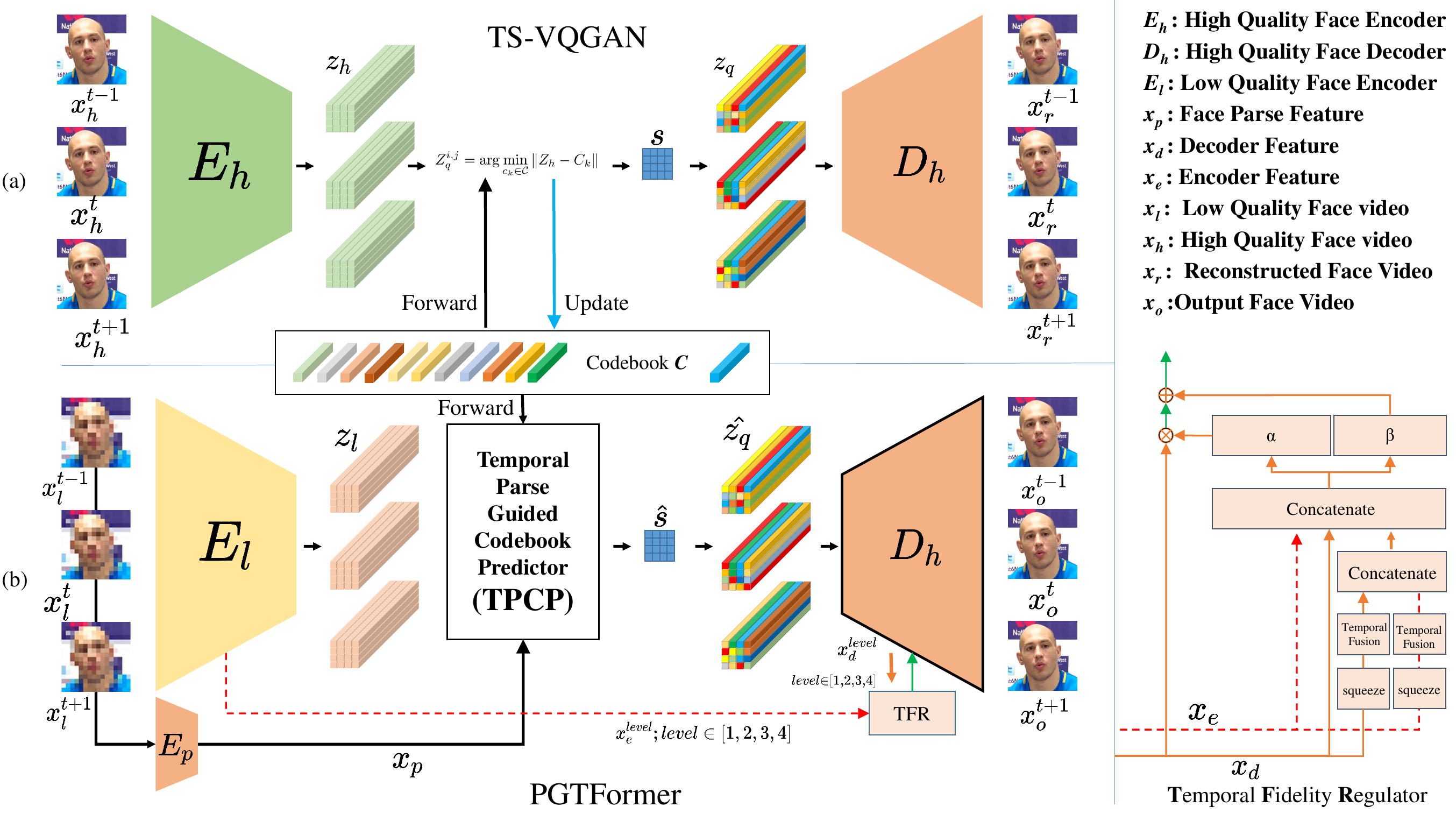} 
\caption{The workflow of the parsing-guided temporal-coherent transformer (PGTFormer). (a) We first learn a temporal-spatial quantized autoregressive encoder (TS-VQGAN) to enable the codebook and decoder to represent high-quality face video sequences. (b) We input the low-quality video sequence into the low-quality face encoder $E_l$ to obtain the low-quality face latent features $z_l$.
Input the low-quality video sequence into the Face Parsing Module $E_p$ to obtain the face parsing features $x_p$.
 Then $x_p$ and $x_l$ are input into the temporal parsing-guided codebook predictor (TPCP) to predict high-quality video face features.
Finally, the low-quality feature $x_e$ is fused with the high-quality feature $x_d$ in face decoder $D_h$. Specifically, we design a temporal fidelity regulator (TFR) to improve the temporal coherence of the face video. 
The weights of $D_h$ are pre-trained in (a) and fixed in (b).
}
\label{ourframework}

\end{figure*}

\section{Related Work}

\subsection{Blind Image Face Restoration}

Current methods based on generator priors can recover high-quality texture details.
For example, \cite{Gu_Shen_Zhou_2020,Menon_Damian_Hu_Ravi_Rudin_2020} use the face generation prior from StyleGAN for blind face restoration, but the degraded LQ face is ill-posed when converting to HQ face, and the fidelity of these methods is difficult to be guaranteed.
In order to solve the problem of too low fidelity, GLEAN\cite{Chan_Wang_Xu_Gu_Loy_2021} and GFPGAN\cite{Wang_Li_Zhang_Shan_2021} directly introduce the generator prior into the decoder, which can improve the fidelity to a certain extent. Despite the increased fidelity, when the input is heavily corrupted, artifact artifacts can be produced.

Another method, VQFR\cite{Gu_Wang_Xie_Dong_Li_Shan_Cheng} and CodeFormer\cite{Zhou_Chan_Li_Loy_2022}, directly uses the codebook dictionary information in the VQGAN series of methods to model face restoration as a codebook selection problem, and improves the fidelity through encoder information fusion.
These methods all use the face prior information of the generator, but highly aligned face data is required to train the generator, and only the aligned face can be restored during use.
If the input face is not pre-aligned, the resulting face image will have unacceptable artifacts.

In the video face, there will be structural changes in the faces of adjacent frames, and this structural change is more prominent in low-quality faces. The difference between two adjacent frames of faces is detected by key points, pre-aligned, face restoration, and objection After aligning the four passes were zoomed in, which caused major problems with discontinuity between frames.

Therefore, although these methods based on generative priors have achieved good performance in image face restoration tasks, they cannot effectively restore low-quality faces in videos.

\subsection{Video Restoration}

Video restoration is to reduce or eliminate the quality loss of video in the process of lossy compression, so as to generate a high-quality image video that is closer to lossless.
MFQE\cite{Guan_Xing_Xu_Yang_Liu_Wang_2021,Yang_Xu_Wang_Li_2018} adds multi-scale information extraction and dense connection mapping structure to the quality enhancement network to improve the enhancement effect. In EDVR\cite{Wang_Chan_Yu_Dong_Loy_2019}, time information is gathered through space-time deformable convolution. According to different image frames, the space-time sampling position of convolution can be deformed adaptively, expanding the receptive field to a certain extent, and obtaining the most relevant context. information, exclude noisy content and improve the quality of image frames. The BasicVSR\cite{Chan_Wang_Yu_Dong_Loy_2021,Chan_Zhou_Xu_Loy_2022} series methods restore high-quality videos through feature propagation and alignment.

\section{Methodology}

The primary objective of our research is to achieve superior restoration of the face in video. However, traditional image-level face restoration methods encounter several limitations when applied to videos. These limitations include the inability to leverage temporal information to enhance restoration quality, the lack of continuity in restored faces, and the introduction of additional degradation due to frame-based pre-alignment.

To address these challenges, we propose a comprehensive framework. The workflow of the framework is illustrated in Fig. \ref{ourframework}. It comprises two main parts. The first part is a spatio-temporal quantized autoencoder (TS-VQGAN) based on Swin3D, as depicted in Fig. \ref{ourframework} (a). This component is trained on high-quality face videos and captures high-quality texture priors.
The second part is the temporal parse-guided codebook selection face restoration network, as shown in Fig. \ref{ourframework} (b). This component utilizes temporal parse information to guide the codebook selection process, thereby facilitating accurate restoration of facial features.

TS-VQGAN allows us to extract facial information from videos and generate a quantized codebook representation of high-quality facial video sequences.
After obtaining a codebook with high-quality video face priors, the face restoration problem is modeled as a codebook prediction problem.
Subsequently, our focus shifts towards designing the corresponding codebook predictor. To accomplish this, we employ a 3D transformer as the codebook selector. This codebook selector utilizes the facial features extracted from the video sequence to accurately select the appropriate high-quality codebook. The selected codebook is then fed into the decoder $D_h$ to recover a high-quality face video sequence. To effectively handle complex facial poses in the video, which may lead to erroneous codebook selection by the codebook selector, we introduce the incorporation of face parse features as conditional position codes. This ensures accurate codebook selection even in the presence of atypical facial poses, thereby avoiding codebook selection errors caused by fixed face space priors. Finally, we introduce a temporal feature regulator to enhance the temporal continuity of the recovered video face.

\subsection{Temporal-Spatial VQGAN}

To obtain a high-quality representation of video faces, we adopt a pre-training strategy using the temporal-spatial quantized autoencoder (TS-VQGAN). This approach aims to learn a codebook that encompasses spatio-temporal context, thereby enhancing the network's expressiveness and robustness against degradation.

As depicted in Fig. \ref{ourframework} (a), the input consists of a sequence of high-quality face video frames denoted as $x_h \in \mathbb{R}^{T \times H \times W \times 3}$. These frames are fed into the spatio-temporal interactive encoder $E_h$, which generates a potential feature representation $z_h \in \mathbb{R}^{T \times H \times W \times D}$. Subsequently, the nearest neighbor matching technique is used to find the closest codebook vector $C_k$ in codebook $C$ for each vector in $z_h$, and the index $S$ of the codebook is obtained. Get the  quantized representation latent $z_q$ through $S$. For the formal expression, please refer to Equation \ref{vqganformal}. Finally, the quantized representation $z_q$ is input into the decoder $D_h$, yielding the reconstructed video sequence $x_r$.

\begin{equation}
\begin{aligned}
z_h &= E_h(x_h;\theta_{E_h}), \\
z_q &=\arg\min_{C_k\in\mathcal{C}}\|z_h-C_k\|_2, \\
x_r &= D_h(z_q;\theta_{D_h}),
\end{aligned}
\label{vqganformal}
\end{equation}
$\theta_{E_h}$, $\theta_{D_h}$ are the parameters of the encoder and decoder, respectively.
Here, our encoder $E_h$ and decoder $D_h$ are comprised of 8 residual blocks and 4 resizing layers for downsampling and upsampling. To ensure accurate face reconstruction, we set the number of codebooks $N$ to 1024, which provides ample capacity. Furthermore, we set the code dimension $d$ to 512. To extract video sequence features effectively, we integrate Swin3D Transformer layers during the 2nd, 3rd, and 4th downsampling and upsampling stages.

\subsection{Temporal Parse-guided Codebook Predictor}

Utilizing the trained TS-VQGAN's codebook and decoder to generate high-quality faces, we formulate the face restoration problem as a controllable face generation process.

During the face restoration process, we input the low-quality face $x_l$ into the low-quality encoder $E_l$, generating a low-quality latent representation $z_l$. Our goal is to use the function $T$ to predict the codebook index $\hat{S}$ based on $z_l$, and obtain the corresponding high-quality features$\hat{z_q}$ from Codebook $C$ according to the index $\hat{S}$. By inputting $\hat{z_q}$ into the decoder $D_h$, we obtain the restored output $x_o$.

This process can be framed as a maximum likelihood estimation, where we maximize the conditional likelihood of the predicted codebook $\hat{z_q}$ with respect to the low-quality input image $x_l$, enabling optimal parameter estimation for $E_l$ and $T$ to achieve accurate code prediction. Due to the loss of face texture details in the low-quality input, the accuracy of codebook selection using the nearest neighbor method is limited. To address this, previous methods employ a simple Transformer to predict a single frame codebook. 

The process can be formally defined as follows.

\begin{equation}
\begin{aligned}
z_l &= E_l(x_l,\theta_{E_l}), \\
\hat{z_q} &= T(z_l; \theta_T), \\
x_{o} &= D_h(\hat{z_q},\theta_{D_h}),
\end{aligned}
\label{ll0}
\end{equation}

\noindent 
where $\theta_{E_l}$, $\theta_T$, and $\theta_{D_h}$ represent the parameters of the encoder, Transformer, and decoder, respectively. By optimizing these parameters, we aim to enhance the face restoration performance and generate visually appealing results.

\begin{figure*}[t]
\centering
\includegraphics[width=0.99\linewidth]{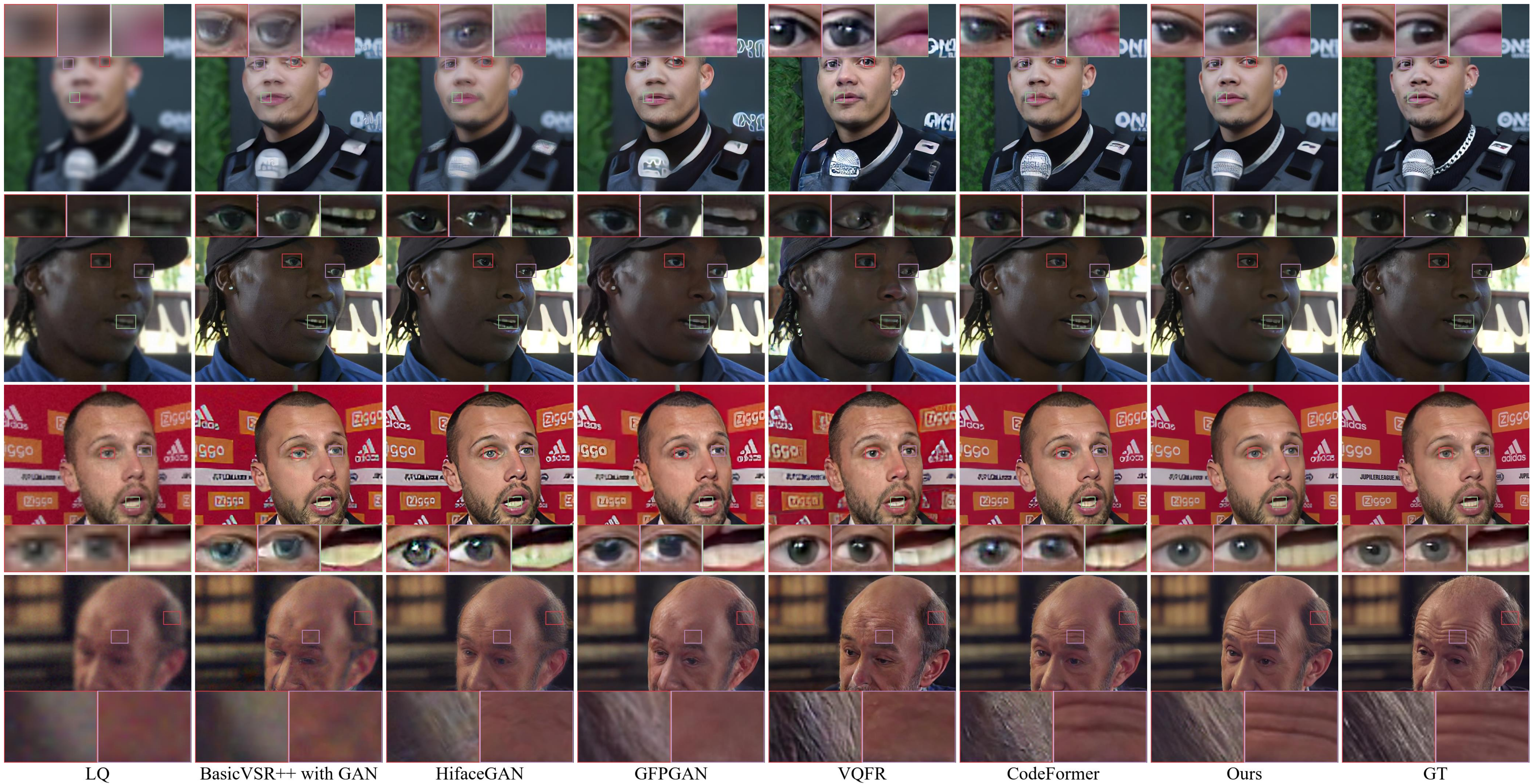} 
\caption{Qualitative comparison. We show the results of aligned and non-aligned face videos. We can see that our method has fewer artifacts and can restore face information more naturally.}
\label{qual0}
\vspace{-0.2cm}
\end{figure*}

\begin{table*}[t]
  \centering
  \caption{Quantitative comparison for both aligned and non-aligned face video training and testing.}
  
 \resizebox{\textwidth}{!}{%
    \begin{tabular}{c|ccccccc|ccccccc}
    \toprule
    & \multicolumn{7}{c|}{Train \& Test on Aligned Face Video} & \multicolumn{7}{c}{ Train \& Test on 
 Non-Aligned Face Video} \\
    \midrule
    Method & PSNR & SSIM & LPIPS & Deg & LMD & TLME & MSRL & PSNR & SSIM & LPIPS & Deg & LMD & TLME & MSRL \\
    \midrule
    BasicVSR++ & 26.70 & 0.7906 & 0.3086 & 38.31 & 2.89 & 6.97 & 24.15 & 26.17 & 0.7482 & 0.3594 & 36.14 & 2.39 & 7.09 & 23.91 \\
    HiFaceGAN & 28.45 & 0.8297 & 0.2924 & 34.02 & 2.25 & 5.73 & 25.81 & 27.41 & 0.7926 & 0.3167 & 32.74 & 1.99 & 5.59 & 24.99 \\
    GFPGAN & 27.77 & 0.8252 & 0.2893 & 31.70 & 2.40 & 6.11 & 25.68 & 26.27 & 0.7864 & 0.3167 & 30.14 & 2.13 & 6.17 & 24.69 \\
    VQFR & 25.59 & 0.7788 & 0.3003 & 37.83 & 2.99 & 7.41 & 23.60 & 25.33 & 0.7459 & 0.3134 & 33.27 & 2.40 & 7.05 & 23.04 \\
    Codeformer & 28.71 & 0.8226 & 0.2460 & 28.11 & 1.97 & 5.82 & 26.32 & 27.88 & 0.8018 & 0.2738 & 26.55 & 1.74 & 5.60 & 25.54 \\
    Ours & \textbf{30.74} & \textbf{0.8668} & \textbf{0.2095} & \textbf{24.41} & \textbf{1.63} & \textbf{4.20} & \textbf{28.16} & \textbf{29.66} & \textbf{0.8408} & \textbf{0.2230} & \textbf{23.09} & \textbf{1.35} & \textbf{4.09} & \textbf{27.33} \\
    \bottomrule
    \end{tabular}%
    }
    \vspace{-0.4cm}
  \label{combined-table}%
\end{table*}%

However, for video face restoration, straightforward using the Transformer prediction codebook cannot effectively utilize temporal context face information, and artifacts are prone to appear on unaligned low-quality video faces (such as the Fig. \ref{parandvqgan}).

Based on this, 
we design a temporal parse-guided Non-Autoregressive Transformer for codebook prediction. This design pattern can predict the codebook more accurately and restore high-quality face videos.
Specifically, we input the video sequence into the face parsing model $E_p$, and embed the obtained intermediate feature $x_p$ into the Transformer as a relative position code.
The network structure of the face parsing model $E_p$ is consistent with \cite{yu2021bisenet}, and we use the CelebAMask-HQ \cite{CelebAMaskHQ} dataset to train $E_p$.
At this time, the process of our temporal parse-guided codebook predictor can be defined in the Eq. (\ref{ll1}).

\begin{equation}
\begin{aligned}
z_l &= E_l(x_l;\theta_{E_l}), \\
x_p &=E_p(x_l;\theta_{E_p}), \\
\hat{z_q} & = TPCP(z_l; x_p; \theta_{TPCP}). \\
\end{aligned}
\label{ll1}
\end{equation}

Compared with previous codebook predictior, our proposed codebook predictor TPCP has two advantages: 1) It can utilize multi-frame information for more accurate codebook prediction. 2) Face parsing is used as position coding to realize codebook selection for face parsing perception.

\subsection{Temporal Fidelity Regulator}

Although it is already possible to obtain high-quality face videos using codebook prediction, the fidelity and coherence of faces are also important when recovering faces from videos.
To this end, this paper designs a Temporal Fidelity Regulator (TFR) that uses the encoder feature $x_e$ to modulate the decoder feature $x_d$. Specifically, our TFR is divided into two processes.
The first is time feature compression, which performs dimensionality reduction (squeeze) on the input $x_e$ and $x_d$. Followed by time feature aggregation, temporal fusion is performed on $x_e$ and $x_d$ along the time dimension to get $ x^{T}_{e}$ and $x^{T}_{d}$ through convolution. Next, concatenate the interacted $x^{T}_{e}$, $x^{T}_{d}$ with $x_e$ and $x_d$ to get $x_{concat}$. After two convolutions respectively, affine parameters $\alpha$ and $\beta$ are obtained, and $\alpha$ and $\beta$ are used to modulate the multi-level decoder feature $x_d$ to obtain the output feature $\hat{x_d}$. The TFR will be adopted in each decoder level to enhance the final temporal refined output $x_o$.  The detail of TFR is shown in Eq.\ref{tfrformula2}.

\begin{equation}
\begin{aligned}
&\quad x^{f}_{e}, x^{f}_{d} = Conv(x_e), Conv(x_d) \\
& x^{T}_{e} = Conv(Reshape(x^{f}_{e}, (b \times t, c, h, w))) \\
& x^{T}_{d} = Conv(Reshape(x^{f}_{d}, (b \times t, c, h, w))) \\
&\quad x_{concat} = Concat(x^{T}_{e}, x^{T}_{d}, x_e, x_d) \\
&\quad \alpha, \beta = Conv(x_{concat}) \\
&\quad \hat{x_d} = \alpha \cdot x_d + \beta \\
\end{aligned}
\label{tfrformula2}
\end{equation}

TFR performs feature aggregation in the time dimension during the modulation process, thereby improving the consistency of restored video faces and reducing jitter artifacts.

\subsection{Training Objectives}

Based on our sophisticated architecture, the training of the model is divided into four stages.

\textbf{Stage I}.
In the training process of the quantized autoencoder, three loss functions ($\mathcal{L}_1$, perceptual loss, adversarial loss) are used to jointly optimize TS-VQGAN, and the specific form is shown in the Eq. \ref{f1}, $\Phi$ is the VGG19 network. The weights of the three losses are set to 1.0, 1.0, and 0.75, respectively.

\begin{eqnarray}
\begin{aligned}
\mathcal{L}_1&=\|x_h-x_{r}\|_1 \\
\quad\mathcal{L}_{per}&=\|\Phi(x_h)-\Phi(x_{r})\|_2 ^2 \\
\quad\mathcal{L}_{adv}&=[\log D(x_h)+\log(1-D(x_{r}))]
\label{f1}
\end{aligned}
\end{eqnarray}

\textbf{Stage II}.
The goal of the second stage is to fine-tune the encoder $E_l$ and train the parsing-guided  codebook predictor (TPCP).
Only code-level loss is needed at this stage.
The codebook prediction loss comprises two distinct components. The primary component, denoted as \( L_{\text{code}} \), involves computing the cross-entropy loss between the predicted and actual codebook indices. The secondary component pertains to the codebook's feature loss, where we measure the L2 loss between the encoder's output feature \( z_l \) and the ground truth codebook feature \( z_q \). 

\begin{eqnarray}
\begin{aligned}
\quad\mathcal{L}_{code}&=-\frac1N\sum_{n=0}^{N-1}s \cdot log(\hat{s}) \\
\quad\mathcal{L}_{z}&=\| z^l - z^q \|_2 ^2,
\label{fs2}
\end{aligned}
\end{eqnarray}

\noindent
where $s$ is the codebook selected by the HQ video input to the HQ encoder, and $\hat{s}$ is the codebook predicted by the LQ video input to the LQ encoder.

\textbf{Stage III}.
The third stage is fidelity training for PGTFormer.
To optimize the TFR module and fine-tune the encoder $E_l$, we use code loss $\mathcal{L}_{code}$, pixel reconstruction loss $\mathcal{L} _1$, perceptual loss $\mathcal{L}_{per}$, adversarial loss $\mathcal{L}_{adv}$, and set the weights to 0.5, 1.0, 1.0, 1.0, and 0.75,  respectively. 

\textbf{Stage IV}.
The fourth stage is complex pose face fine-tuning training, here we use unaligned face video sequences to fine-tune the model, and the loss is consistent with the third stage. 
We added $\mathcal{L}_{MSRL}$\cite{dai2022video} as a loss to increase video coherence, defined in Eq. (\ref{msrlloss}).
\begin{eqnarray}
\begin{aligned}
\mathcal{L}_{MSRL}&=\frac{1}{3} \sum_{n=1}^{3} \|(({x^{t+1}_h})\downarrow_{n} - (x^{t}_h)\downarrow_{n}) - \\
&((x^{t+1}_{o})\downarrow_{n} - (x^{t}_{o})\downarrow_{n})\|_1
\label{msrlloss}
\end{aligned}
\end{eqnarray}

\section{Experiments}

\subsection{Datasets and Metrics}

\textbf{Datasets}.
We train the model on the VFHQ \cite{xie2022vfhq} dataset, which contains 16,000 video sequences, the resolution of the video is 512 $\times $ 512. To make training pairs, we follow \cite{xie2022vfhq}  synthesize LQ videos $x$ from HQ video counterparts $y$ with the following degradation model (in Eq.\ref{dde}):

\begin{equation}
x = [(y \circledast k_{\sigma}) {\downarrow}_r + n_{\delta}]_{ \mathrm{FFMPEG}_{crf}}
\label{dde}
\end{equation}

\noindent 
where the HQ video $y$ is first convolved with a Gaussian kernel $k_\sigma)$, and then downsampled to scale $r$. After that, additive Gaussian noise $n_\delta$ is added to the video, 
and then the video coding degradation is implemented by FFMPEG(x264\cite{merritt2006x264}) with a constant rate factor (CRF). 
Finally, the LQ video is resized to 512×512. The procedure for generating complex video bind face degradation is provided by VFHQ.

\textbf{Metrics}.
We first use PSNR, SSIM, LPIPS \cite{Zhang_Isola_Efros_Shechtman_Wang_2018} to evaluate the quality of generated faces. In addition, we also evaluate identity preservation by the cosine similarity of the features of ArcFace \cite{Deng_Guo_Yang_Xue_Cotsia_Zafeiriou_2021,sun2020face}, denote as Deg. To better measure the fidelity of accurate facial expressions and expressions, we further employ Landmark Distance (LMD) as a fidelity metric. Specifically, we use the method proposed by \cite{Wang_Bo_Fuxin_2019} to obtain 98 landmarks from recovered faces and real faces. We then compute the L2 distance for each landmark and average the distances into the final score for the LMD metric.
Since we are a video face restoration task, it is necessary to evaluate the continuity of the face, we calculate the L2 distance between the real face landmark displacement and the recovered face landmark displacement, which can evaluate the continuity of the face components in the video.
We further calculate the peak signal-to-noise ratio of MSRL \cite{dai2022video} to evaluate the continuity of the restored face video.

\begin{table}[h]
  \centering
  \caption{Comparison of inference time and FVD results (Frechet Video Distance), previous methods require preprocessing (landmark detection, warping, inverse warping).}
  
 \resizebox{0.48\textwidth}{!}{%
    \begin{tabular}{cccccc}
    \toprule
    Method & BasicVSR++ & GFPGAN & VQFR  & Codeformer & Ours \\
    \midrule
    Runtime(s) & 0.2388 & 0.1704 & 0.2324 & 0.1217 & \textbf{0.1121} \\
    \midrule
    FVD $\downarrow$ & 467.9 & 326.75 & 425.85 & 217.51 & \textbf{165.19} \\
    \bottomrule
    \end{tabular}%
    }
  \label{ablruntime}%
\end{table}%

\subsection{Comparisons with State-of-the-Art Methods}
\subsubsection{Quantitative Results.} We compare the proposed PGTFormer with state-of-the-art methods, including HiFaceGAN, GFPGAN, VQFR, CodeFormer.
We also compare with the current state-of-the-art method BasicVSR++ for video restoration.
To verify the effectiveness of different methods, we train and test models on aligned and unaligned data.

\begin{figure}[h]
    \centering
    \includegraphics[width=0.98\linewidth]{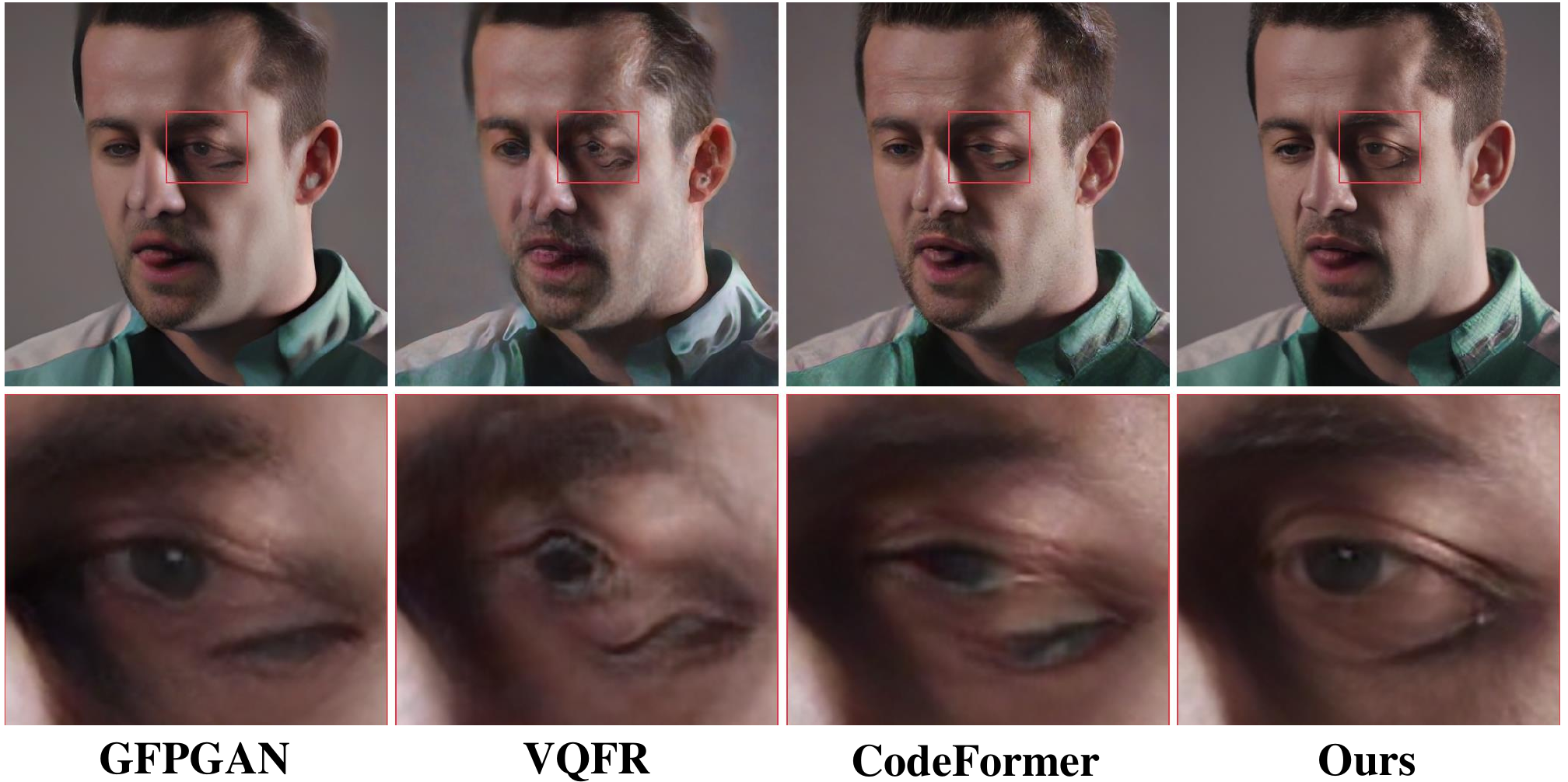}
    \caption{Previous methods generate artifacts when input with a non-standard pose face.}
    \label{artifactsimg}
\end{figure}

\begin{table}[h]
  \centering
  \caption{Ablation of TS-VQGAN. TS-VQGAN is better than VQGAN on multiple evaluation metrics because TS-VQGAN uses video information to help better represent face priors.
}
    \begin{tabular}{cccc}
    \toprule
    Method & PSNR$\uparrow$  & SSIM$\uparrow$  & LPIPS$\downarrow$ \\
    \midrule
    VQGAN & 27.65 & 0.844 & 0.1813 \\
    Temporal-VQGAN & 28.85 & 0.8614 & 0.1466 \\
    \bottomrule
    \end{tabular}%
  \label{abltabletvqgan}%
\end{table}%

\begin{figure}[h]
\centering
\includegraphics[width=0.99\linewidth]{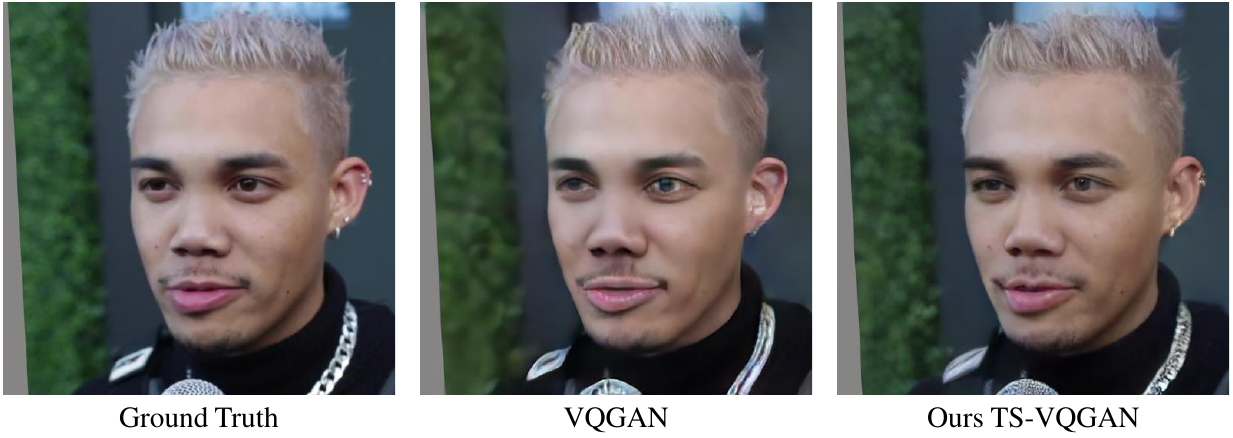} 
\caption{Ablation of TS-VQGAN.The high-quality face prior embedded by TS-VQGAN has fewer artifacts and higher fidelity.}
\label{abltvqgan}
\end{figure}

We show the quantitative comparison of VFHQ-Test in Table \ref{combined-table}(tests on aligned and unaligned). In terms of image quality metrics PSNR, SSIM, LPIPS, our method achieves the best scores than existing methods. Furthermore, it faithfully preserves the identity, which is reflected in the highest Deg score. We calculated the landmark distance LMD between the restored face and the ground truth, which can evaluate the structural difference of the face, and our method achieved best performance on the LMD index.
We evaluate the temporal continuity by tracking the offset error of the face landmarks between frames, which can evaluate the temporal continuity of different methods.
In Table \ref{combined-table}, our method achieves the best performance on TLME and MSRL metrics, proving that our method has higher video continuity and can restore more natural continuous face video.

In addition, we calculated the FVD metric, which can evaluate the subjective quality of the recovered face video. Table \ref{ablruntime} shows the FVD score. Compared with the past state-of-the-art methods, our proposed method has significant advantages in the FVD metric.

\subsubsection{Qualitative Results.}

Furthermore, we give a qualitative comparison in Fig. \ref{qual0}. Other comparison methods did not yield pleasant recovery results.
Especially in the restoration quality of eyes and teeth, other methods are prone to disturbing hallucinatory artifacts, which our method can overcome.
Since our face prior comes from spatio-temporal interaction features, it can better characterize face components. In areas prone to artifacts such as eyes and teeth, our method can recover natural face components. The advantages of our method in the quality of the eye region are demonstrated in Fig.\ref{artifactsimg}.

\subsection{Ablation Studies}

\subsubsection{Importance of TS-VQGAN.}
We do ablation experiments to study the superiority of TS-VQGAN. As shown in Fig.\ref{abltvqgan}.
The left side is the ordinary VQGAN reconstruction result, and the right side is the reconstruction result of our proposed TS-VQGAN. We can see that our Temporal has higher fidelity and texture details. This is because in the process of expressing facial features, spatio-temporal feature interaction can further complement the missing details of a single frame, and implicitly align features through Swin3D. Quantified ablation results such as Table \ref{abltabletvqgan}, compared with ordinary VQGAN, we have advantages in evaluation indicators such as PSNR, SSIM, and Lpips. This proves that our TS-VQGAN can better reconstruct video faces.

\begin{table}[t]
  \centering
  \caption{Ablation for Parsing-Guided Codebook Predictor(Stage II). Baseline, Face parsing guide, and 3D codebook predictor are abbreviated as B, M0, and M1 respectively.}
    \begin{tabular}{cccccccc}
    \toprule
    B & M0 & M1 & PSNR $\uparrow$  & LPIPS$\downarrow$ & Acc\_T1 & Acc\_T5\\
    \midrule
    \checkmark     &                &                & 26.55 & 0.2494 & 20.49 & 44.68 \\
    \checkmark     & \checkmark     &                & 26.60 & 0.2485 & 20.78 & 45.05 \\
    \checkmark     & \checkmark     & \checkmark     & 26.63 & 0.2475 & 20.91 & 45.23 \\
    \bottomrule
    \end{tabular}%
  \label{ablpg3d}%
\end{table}%

\begin{table}[h]
  \centering
  \caption{The ablation result of temporal fidelity regulator(TFR). The introduction of the TFR module can enhance the continuity of facial landmarks in Temporal.}
    \begin{tabular}{cccccc}
    \toprule
    Baseline & TFR   & Deg $\downarrow$  & LMD  $\downarrow$ & TLME $\downarrow$ & MSRL $\uparrow$\\
    \midrule
      \checkmark&                 & 25.70  & 1.52  & 4.38 & 27.23 \\
      \checkmark&       \checkmark& 23.09 & 1.35  & 4.09 & 27.33 \\
    \bottomrule
    \end{tabular}%
  \label{ablttabletfr}%
\end{table}%

\begin{table}[h]
  \centering
  \caption{Ablation of Parsing-Guided Codebook Predictor (Stage III). The face parsing prior and the 3D codebook predictor significantly improve the face restoration quality.}
    \begin{tabular}{lccc}
    \toprule
    \multicolumn{1}{c}{Module} & \multicolumn{1}{c}{PSNR} $\uparrow$ & \multicolumn{1}{c}{SSIM} $\uparrow$ & \multicolumn{1}{c}{LPIPS} $\downarrow$ \\
    \midrule
    Baseline & 29.20 & 0.8379 & 0.2245 \\
    + Face Parse Prior & 29.79 & 0.8469 & 0.2196 \\
    + 3D Codebook Predictor & 30.82 & 0.8674 & 0.2087 \\
    \bottomrule
    \end{tabular}%
  \label{tab4}%
\end{table}%

\subsubsection{Superiority of Temporal Parse-guided Codebook Predictor (TPCP).}
In order to ablate the effectiveness of the proposed analytically guided 3D codebook predictor, we add analytically guided and 3D codebook predictors on the basis of Baseline respectively. From the Table \ref{ablpg3d} and Table \ref{tab4} we can see that with the introduction of face parse guidance and temporal  codebook predictor, indicators such as PSNR have been improved. In particular, we have further measured the accuracy of codebook selection, and we can see that the introduction of the two components has improved Top 1 Acc and Top 5 Acc.

\subsubsection{Effectiveness of Temporal Fidelity Regulator (TFR)}. 
We ablate temporal fidelity regulator (TFR), and show the results of ablation experiments in Table \ref{ablttabletfr}. After adding TFR, Landmark Continuity Error (TLME) is reduced and Multiscale Continuity (MSRL) is improved.
It is proved that the proposed TFR can utilize the temporal features to enhance the continuity of the restored video face.

\subsubsection{Inference Time.}
We analyze the inference speed of different methods. The previous method applied to the video face restoration task requires additional pre-alignment operations. We calculated the total inference speed (keypoint detection, warp, restoration, inverse warp), in the table The time overhead of different methods is shown in Table \ref{ablruntime}.
Compared with previous face restoration method our method achieves the best performance on inference time.

\section{Conclusion}

This paper tackles the fundamental challenge of blind video-based face restoration, enabling high-quality results without pre-alignment. To enhance video face representation, we propose TS-VQGAN, which achieves higher-quality face codebook representation. Our design includes a temporal parse-guide codebook predictor that accurately selects the codebook from the continuous video sequence representation. Additionally, we leverage the temporal fidelity regulator to enhance the overall continuity of video face restoration. Notably, our approach is the first to specifically target video face restoration, and our experimental results showcase its superior effectiveness.

\section*{Acknowledgments}
This Research is Supported by the National Key Research and Development Program from the Ministry of Science and Technology of the PRC (No.2021ZD0110600), Sichuan Science and Technology Program (No.2022ZYD0116), Sichuan Provincial M. C. Integration Office Program, and IEDA Laboratory of SWUST, Grant CEIEC-2022-ZM02-0247.

\bibliographystyle{named}
\bibliography{ijcai24}

\end{document}